\documentclass{ws-mpla}

\usepackage{subcaption}
\usepackage[super]{cite}
\usepackage{xcolor}
\usepackage[verbose,hypertexnames=false]{hyperref}
\usepackage{lipsum} 
\usepackage{color}
\definecolor{blue}{rgb}{0.0, 0.0, 1.0}
\definecolor{red}{rgb}{1.0, 0.0, 0.0}
\definecolor{royalblue}{rgb}{0.0, 0.14, 0.4}
\hypersetup{colorlinks=true,linkcolor=red, citecolor=blue, urlcolor=blue}
\usepackage{graphicx}
\usepackage{dcolumn}
\usepackage{bm}
\usepackage{amsmath,amssymb}
\usepackage{float}
\usepackage{multirow}
\usepackage{slashed}
\usepackage{xcolor}
\usepackage{braket}
\usepackage{physics}
\usepackage{multirow}
\usepackage{gensymb}
\def\orcid#1{\kern .08em\href{https://orcid.org/#1}{\includegraphics[keepaspectratio,width=0.7em]{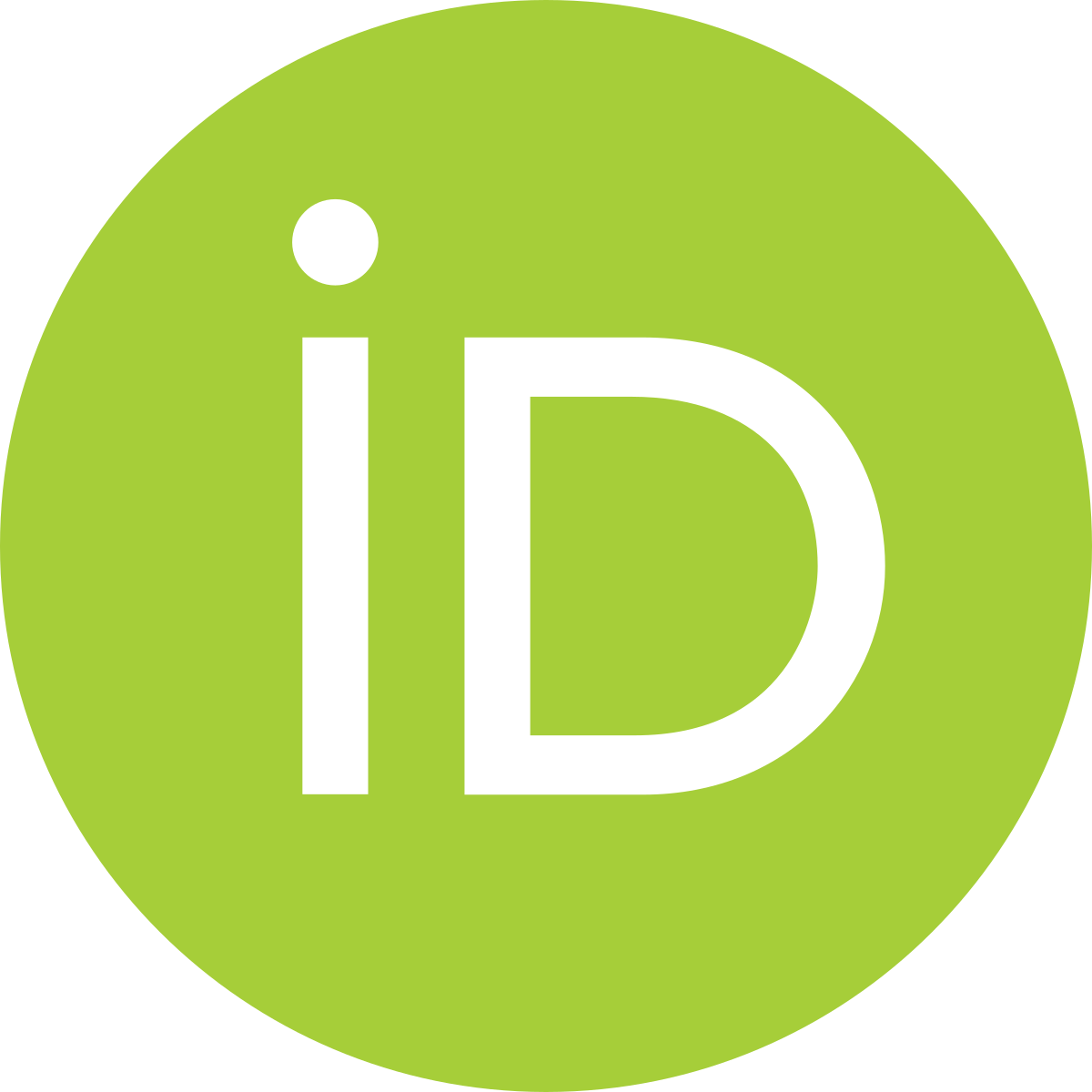}}}

\begin{document}

\thispagestyle{plain}

\title{A Brief History and Outlook of Hadronic Physics in Indonesia}

\author{Ahmad Jafar Arifi\orcid{0000-0002-9530-8993}}
\address{Advanced Science Research Center, Japan Atomic Energy Agency, Ibaraki 319-1195, Japan \\
Research Center for Nuclear Physics, The University of Osaka, Ibaraki, Osaka 567-0047, Japan \\
E-Mail Address: aj.arifi01@gmail.com}

\author{Parada~T.~P.~Hutauruk\orcid{0000-0002-4225-7109}}
\address{Department of Physics, Pukyong National University (PKNU), Busan 48513, Korea\\
Departemen Fisika, FMIPA, Universitas Indonesia, Depok 16424, Indonesia\\
E-Mail Address: phutauruk@gmail.com}
\author{Terry Mart\orcid{0000-0003-4628-2245}}
\address{Departemen Fisika, FMIPA, Universitas Indonesia, Depok 16424, Indonesia\\
E-Mail Address: terry.mart@sci.ui.ac.id}
\author{Chalis Setyadi\orcid{0000-0002-5853-4238}}
\address{Department of Physics, Universitas Gadjah Mada, BLS 21, Yogyakarta, Indonesia \\
E-Mail Address: chalis@mail.ugm.ac.id}


\maketitle

\begin{abstract}
Hadronic physics has gradually emerged as one of the growing research frontiers in Indonesia, driven by efforts to better understand the properties of the strong interaction and the internal structure of hadrons from the fundamental principles of Quantum Chromodynamics. In the last few decades, Indonesian researchers have made significant contributions to developing various theoretical and phenomenological aspects of hadrons. 
In addition, on the experimental side, Indonesian scientists have participated in hadron experiment facilities overseas, such as the ALICE collaboration at CERN, which has strengthened the scientific activities and networks and has supported the training of young Indonesian researchers. In the present paper, we review Indonesian scientists’ contributions to hadronic physics, highlight ongoing research directions in both experimental and theoretical, and outline strategies for future development toward integration into the international hadronic physics community.
\end{abstract}

\section{Hadronic Physics and QCD}

Hadronic physics serves as a bridge between the microscopic world of quarks and gluons and the macroscopic properties of nuclear matter, linking fundamental QCD theory~\cite{Gross:2022hyw} with nuclear structure, astrophysics, and the dynamics of the early universe. Through continued theoretical, computational, astrophysical, and experimental efforts, the field strives to uncover how the strong force gives rise to the visible matter that constitutes the universe. This aligns with the Topical Group on Hadronic Physics in the American Physical Society (APS) unit\footnote{\url{https://www.aps.org/membership/units}} announced in 2002, which recognized hadronic physics as a distinct and essential component of the APS research community. According to its Bylaws, the group’s objective is to advance and disseminate knowledge related to the physics of the strong interaction. While promoting interdisciplinary research that connects high-energy, nuclear, and condensed-matter physics, its primary scientific focus includes, but is not limited to, hadron spectroscopy and structure, lattice gauge theory, heavy-ion physics, jet physics, QCD at high energies, and QCD at finite temperature and density. A dedicated topical section on ``Hadronic Physics and QCD'' has likewise been featured in Physical Review C.

The following sections provide an overview of the historical development, research activities, and community efforts in hadronic physics in Indonesia, followed by a discussion of future perspectives and challenges likely to shape the growth of this field in the coming years.
It is worth noting that this article is not intended to create an exclusive community but rather to identify the researchers involved in hadronic physics and their activities, which cover a relatively small segment of the broader high-energy physics (HEP) community in Indonesia, and we hope this can stimulate similar reviews in other HEP subfields. Additionally, we anticipate that this brief review will provide international colleagues and communities with a clearer understanding of the current physics landscape and the individuals (theoretical and experimental physicists) involved in this field.

\section{Historical Hadronic Physics Development}
The development of hadronic physics in Indonesia reflects a gradual evolution of research capacity, human resources, and the broader scientific community. 
Viewed across these years of development, the growth of hadronic physics in Indonesia can be broadly divided into several stages as follows:

\subsection{Pioneering Stage (1970s--2000s)}
In the early period, hadronic physics in Indonesia had limited exposure and was only beginning to develop. Meanwhile, theoretical physics was more widely developed and recognized. 
While many individuals contributed to establishing physics departments across the country, our emphasis here is on those who conducted hadronic physics research, as reflected in their past and present research records.

One of the earliest Indonesian hadron physicists, M. Barmawi~\cite{Barmawi:1966zz,Barmawi:1968zz}, completed his Ph.D at the University of Chicago (1964), studying the Regge model for vector meson production under the supervision of Reinhard Oehme. He then worked at the International Centre for Theoretical Physics (ICTP) on a quark model for the pomeron~\cite{Barmawi:1974yr}. Upon returning to Indonesia, he joined a faculty position at Institut Teknologi Bandung (ITB) and later shifted his research focus to material physics. Similarly, Tjia May On, another early Indonesian hadron physicist, earned his doctorate at Northwestern University (1969) under the supervision of Carl H. Albright and Lianshou Liu, focusing on quark-model studies of semileptonic baryon transitions~\cite{Albright:1966uhq}. After further training at ICTP~\cite{Albright:1974nd}, he returned to ITB and eventually moved into material physics. Although both later transitioned to other research fields, they are still recognized as pioneering figures in Indonesian hadronic physics.

Darmadi Kusno~\cite{Kusno:1978is}, a student of Parangtopo,\footnote{Parangtopo was one of the early prominent figures in the Department of Physics at Universitas Indonesia, serving as the department’s chair from 1964 to 1984.} began his career at Universitas Indonesia (UI) before taking a leave to pursue his doctoral studies at the University of Oregon (1979) under the supervision of Michael J. Moravcsik, where he studied how photons interact with nucleons inside the deuteron. He analyzed the effects of nucleon Fermi motion in a covariant framework~\cite{Kusno:1978is}, showing that certain corrections vanish in high-energy hadronic scattering but remain in deep-inelastic electron scattering. After gaining additional experience at the ICTP~\cite{Kusno:1980tf}, he returned to UI.
Next, this group played a pioneering role in establishing hadronic physics in Indonesia. 

Yohanes Surya, a student of Darmadi Kusno, pursued his doctoral studies at the College of William and Mary (1994) with Franz Gross~\cite{Gross:1992tj}, M. T. Pe\~{n}a~\cite{Pena:1996tf}, and Chueng-Ryong Ji~\cite{Ji:1992xr}. His notable contribution to hadronic physics includes the development of a unitary, relativistic resonance model for $\pi N$ scattering~\cite{Gross:1992tj}, which advanced the theoretical understanding of baryon resonances and meson-baryon interactions. 
Yohanes Surya became a key figure in advancing science education in Indonesia. He founded STKIP Surya (later Surya University) to train future scientists and teachers, and through the Indonesian Physics Olympiad, he mentored students who earned numerous medals at the International Physics Olympiad.

Terry Mart~\cite{Mart:1995wu}, a student of Darmadi Kusno, completed his doctoral research on electromagnetic kaon production on the nucleon and nuclei at the Institut f\"ur Kernphysik, University of Mainz (1996), under the supervision of Dieter Drechsel and Cornelius Bennhold. After returning to UI as a faculty member, he helped establish the first substantial hadronic physics research group in Indonesia. The group grew steadily and eventually mentored more than 50 students, most of whom were bachelor’s or master’s students who later pursued graduate studies abroad, either in hadronic physics or in other fields. Meanwhile, the Ph.D. program in theoretical physics at UI was only formally established around 2017, with its first graduate completing the program in 2019, coinciding with the introduction of several double-degree programs. Since then, this group has remained the leading and most active hadronic physics research group in Indonesia.

It is also worth mentioning that L. T. Handoko~\cite{Ali:1996bm} specialized in theoretical physics, particularly in particle physics phenomenology, during his Ph.D studies at Hiroshima University (1998) with Taizo Muta and Takuya Morozumi, and later he went to ICTP and DESY as a postdoctoral researcher. 
Early in his career, he joined the Indonesian Institute of Sciences (LIPI) as a permanent researcher, while also becoming a member of Terry Mart’s group in the Department of Physics at Universitas Indonesia and serving as a part-time lecturer. A decade later, he was elected Chairman of LIPI and subsequently appointed Chairman of the National Research and Innovation Agency (Badan Riset dan Inovasi Nasional, BRIN). Established in 2019, BRIN was created to centralize and strengthen Indonesia’s national research and innovation activities.

In the early stages, alongside hadronic physics, several researchers in Indonesia worked on theoretical and phenomenological particle physics, especially in high-energy physics.
Pantur Silaban went to Syracuse University (1971)~\cite{Silaban:1971} to study general relativity, and upon returning to Indonesia, he became a professor at ITB, widely recognized as the first in the country to pioneer the study of general relativity. He is also well known among undergraduate students for his fundamental physics textbooks, which have served as key references in physics education across Indonesia.
Anggraita Pramudita~\cite{Ma:1981eg} studied at the University of Hawaii in 1981 and later at the ICTP, focusing on particle phenomenology. After returning to Indonesia, he joined BATAN, where he played a central role in the development of the national nuclear energy program.
Hans Jacobus Wospakrik was a faculty member at ITB~\cite{Wospakrik:1982fd} and earned his Ph.D from Durham University (2002) under the supervision of Wojtek J. Zakrzewski, focusing primarily on mathematical physics and soliton (Skyrme) theory.
Freddy Permana Zen earned his Ph.D at Hiroshima University (1994), worked on general relativity, and later joined ITB’s faculty.

At this stage, most of these pioneering physicists were faculty members in Indonesian universities and then pursued doctoral studies in the United States (US), Germany, the United Kingdom (UK), and Japan, reflecting the strong academic connections between Indonesia and those countries. 
In addition, the ICTP played a significant role in the early development of Indonesian hadronic physics.

\subsection{Early Millennium (2000s--2010s)}
During the early 2000s, hadronic physics in Indonesia began to develop and expand, with a new generation of researchers contributing to both theoretical and experimental physics studies.

Anto Sulaksono completed his doctoral studies under the supervision of W. Greiner at the University of Frankfurt (2002)~\cite{Sulaksono:2003re} and is currently a faculty member (professor) at UI. 
His research primarily focuses on nuclear structure, superheavy elements, investigated using the non-relativistic and relativistic mean-field theory, and their applications to neutron stars. Since neutron stars are composed of nucleons, the structure of the nucleon also forms an important part of his scientific interests~\cite{Mart:2013gfa,Suparti:2017msx}. Other notable contributors include Imam Fachruddin, who earned his Ph.D at Ruhr University (2003) under the supervision of Walter Glöckle~\cite{Fachruddin:2000wv}, specializing in few-body nuclear structure, and Agus Salam, a former student of Terry Mart, who completed his doctoral studies at the University of Mainz (2004) with Hartmuth Arenhövel~\cite{Salam:2004gz}, focusing on kaon photoproduction on the deuteron.
Their doctoral studies were largely supported by German DAAD scholarships, and their return to UI represents a significant increase in faculty expertise in hadronic physics.

Suharyo Sumowidagdo, a former undergraduate student of Terry Mart, pursued his Ph.D in experimental particle physics at Florida State University (2008), where he joined the D0 Collaboration~\cite{D0:2004ruf} to study top-quark production. He later became part of the CMS Collaboration, which discovered the Higgs boson in 2012~\cite{CMS:2012qbp}, and subsequently contributed to the ALICE group~\cite{ALICE:2022wpn} at the LHC as a researcher at BRIN. One of the purposes of the ALICE collaboration is to study quark–gluon plasma, a state of matter in which quarks and gluons are deconfined, created in high-energy heavy-ion collisions. Joining the ALICE collaboration marks the beginning of active participation in experimental HEP physics by Indonesian researchers.

Alvin Kiswandhi trained at Florida State University (2006) with Simon Capstick~\cite{Kiswandhi:2003ca}, a leading expert in constituent quark models.
He continued his research at National Taiwan University, collaborating with Ju-Jun Xie and Shin Nan Yang~\cite{Kiswandhi:2011cq}. 
Herry Kwee completed his doctoral degree at the College of William and Mary (2006) under the supervision of Christopher D. Carone, working on spectroscopy, production, and decays of pentaquarks~\cite{Carlson:2003xb}. His brother, Hendra Kwee, also finished his doctoral degree at the College of William and Mary (2008) under the supervision of Henry Krakauer and Shiwei Zhang, working on the correction of finite-size errors in many-body electronic structure calculations~\cite{Kwee_2008}. 

Zainul Abidin studied hadron structure using holographic methods with Carl E. Carlson at the College of William and Mary (2010)~\cite{Abidin:2009hr}, and later joined STKIP Surya, where he focused on physics education. One of his influential works is about the gravitational form factors of the nucleon from a holographic approach~\cite{Abidin:2009hr}, which has recently become a trending topic and a key physics focus of future EIC experiments in the US. His last research work on hadronic structure is the kaon form factor in Holographic QCD collaboration with Parada T. P. Hutauruk~\cite{Abidin:2019xwu}.
Jong Anly Tan obtained his doctoral degree from the College of William and Mary (2010) under the supervision of Joshua Erlich, working on extra dimensions and electroweak symmetry breaking~\cite{Carone:2006wj}.
Ardian N. Atmadja investigated quark–gluon plasma using the AdS/CFT approach~\cite{Atmaja:2010uu} at Leiden University (2010) and subsequently joined BRIN. Before that, he also joined a postgraduate diploma program at ICTP.
His research lies in theoretical and mathematical physics, particularly in holographic methods that connect strongly coupled quantum systems to gravity. 

Besides hadronic physics, it is also worth mentioning several faculty members who work in theoretical, phenomenological, and mathematical physics.
Mirza Satriawan studied generalized parastatistical systems at the University of Illinois at Chicago (2002) and later became a faculty member at Universitas Gadjah Mada (UGM).
Agus Purwanto pursued his doctoral studies in neutrino physics at Hiroshima University (2002) and later joined the Institut Teknologi Sepuluh Nopember (ITS).
He is also well known for his popular discussions on physics and religion, bridging scientific concepts with broader philosophical and cultural topics.
Bobby Eka Gunara earned his Ph.D at Martin Luther University (2003) and later joined the ITB, where he researches general relativity.
Tasrief Surungan obtained his Ph.D from Tokyo Metropolitan University (2004), focusing on phase-transition phenomena, and later became a faculty member at Hasanuddin University.
Husin Alatas earned his Ph.D at ITB (2005) and later became a professor at IPB University, actively working on cosmology and complex systems.

Many other researchers in this time period have also contributed to the development of theoretical physics in Indonesia, though it is not possible to mention them individually in this short review. In the following sections, this review will focus on hadronic physics, as general relativity, particle physics, and other areas have developed along different paths and now represent some of the largest communities within Indonesian theoretical physics. 
By the end of this period, hadronic physics research in Indonesia had become primarily centered at BRIN (LIPI at that time) and UI, forming the core of the national hadronic physics community. 
Other universities, such as ITB, UGM, among others, have also established theoretical physics groups, but with different research focuses.

\subsection{Recent Developments (2010s--present)}
Over the last decade, more Indonesian researchers who have been trained internationally have played a pivotal role in strengthening the national hadronic physics community.

Handhika Satrio Ramadhan completed his undergraduate studies under the supervision of Terry Mart before pursuing a Ph.D at Tufts University (2011), where he specialized in cosmology and gravitational theory. He later joined the faculty at UI and has recently been actively collaborating with Anto Sulaksono on research related to compact stars and black holes~\cite{AlGhifari:2025xrt}. Anton Wiranata
who finished his undergraduate studies under the supervision of Terry Mart and L. T. Handoko, pursued his doctoral degree at Ohio University with Maddapa Prakash and worked on transport coefficients of interacting hadrons~\cite{Wiranata:2011kwb}. Heribertus Bayu Hartanto obtained his doctoral degree under Laura Riena at FSU (2013), where he worked on high-energy QCD processes~\cite{Hartanto:2013aha}. He then pursued postdoctoral research at Aachen University, Durham University, and Cambridge University. Following these appointments, he has recently become a junior group leader at the Asia Pacific Center for Theoretical Physics (APCTP). One of his influential works is about two-loop five gluon scattering in QCD~\cite{Badger:2017jhb}. 
Freddy Simanjutak finished an undergraduate degree under the supervision of Terry Mart and L. T. Handoko at UI, who was a doctoral student of Choong Sun Kim at Yonsei University, worked on charmless decay in a perturbative QCD approach~\cite{Kim:2013cpa}. Nowo Riveli finished his doctoral degree at Ohio University (2014) under the supervision of Justin Frantz and being a part of the PHENIX collaboration, and has worked on the experiment of the direct photon-hadron correlations measurement in Au+Au collision at nucleon center-of-mass energy of 200 GeV with isolation cut methods~\cite{Riveli:2014fpb}. Now, he joins a faculty member at Universitas Padjajaran (UNPAD).

Parada T. P. Hutauruk~\cite{Hutauruk:2016sug,Hutauruk:2018zfk} 
pursued his studies at 
Adelaide University (2016), working on the nonperturbative QCD aspects of the hadron structure under the supervision of Anthony W. Thomas, Young Ross, and Ian Cl\"oet. Next, he continued his research on hadron structure in free space and medium (nuclei)~\cite{Hutauruk:2018cgu} and its applications in the neutrino interaction with matter in a neutron star~\cite{Hutauruk:2018qku} at APCTP and is now a research professor at Pukyong National University.
Ahmad Jafar Arifi pursued his Ph.D at Osaka University, joining the group of Atsushi Hosaka~\cite{Arifi:2020yfp} to study heavy baryon spectroscopy. 
He later moved to APCTP, where he collaborated with Yongseok Oh, Ho-Meoyng Choi, and Chueng-Ryong Ji on meson structure in the light-front model~\cite{Arifi:2022pal}, and with Parada T. P. Hutauruk and Kazuo Tsushima on in-medium modifications of mesons~\cite{Arifi:2023jfe}.
He then joined the Hiyama group at RIKEN before moving to the theory group at the Japan Atomic Energy Agency.
Chalis Setyadi, who completed his undergraduate and master’s degrees at UGM under the supervision of Mirza Satriawan, studied high-energy hadron processes with Dani\"{e}l Boer for his Ph.D at the University of Groningen (2023)~\cite{Boer:2021upt,Boer:2023mip}
and now serves as a faculty member at UGM. His research focuses primarily on gluon generalized transverse-momentum dependent distributions (GTMDs) and high-energy vector meson production, which are relevant to the future EIC experiment. Another bachelor student of Mirza Satriawan at UGM, Khoirul Faiq Muzakka, continued his Master’s studies at the Ludwig-Maximilian University of Munich and the Technical University of Munich in Germany, and later pursued a Ph.D at the University of M\"{u}nster, where he researched global nuclear parton distribution functions (PDFs) using neutrino deep inelastic scattering (DIS) and LHC data \cite{Muzakka:2022wey}.

Hadronic experimentalist Zulkaida Akbar, trained at FSU (2018). During his doctoral degree, he participated in the CLAS collaborations at JLab~\cite{CLAS:2017yjv}
and later joined the SpinQuest experiment at Fermilab in his postdoctoral position at the University of Virginia, focusing on the search for nucleon resonances and the study of the internal structure of nucleons, in particular, the sea quark distribution inside nucleons. M. Jauhar Khalili went for a doctoral degree at KEK Tsukuba (2019), developing the particle detectors and their electronic system.
Syaefudin Jaelani went to Utrecht University (2021) to work with Alexandro Grelli and was involved in the ALICE collaboration~\cite{ALICE:2017thy}. Now, all of them help establish an experimental HEP group at BRIN, along with Suharyo Sumowidagdo. 
Moreover, Catur Wibisono completed a Ph.D degree in experimental nuclear physics at FSU (2025)~\cite{Gray:2024cof} and is currently working at BRIN, while Nizar Septian is currently pursuing a Ph.D degree in experimental hadronic physics at FSU~\cite{GlueX:2025hve}. 

Jason Kristiano completed his undergraduate thesis (2018) under the supervision of Terry Mart, focusing on the development of a pure spin-3/2 propagator for applications in particle and nuclear physics, particularly in kaon photoproduction~\cite{Kristiano:2017qjq}.
After finishing his undergraduate study, he pursued his master’s and doctoral degrees at the University of Tokyo (2024), conducting research on primordial black hole formation under the guidance of Jun'ichi Yokoyama~\cite{Kristiano:2022maq}.
He is currently a postdoctoral researcher at the Yukawa Institute for Theoretical Physics, Kyoto University. 
Having completed his bachelor’s and master’s degrees under the supervision of Terry Mart, Samson Clymton~\cite{Clymton:2021wof} pursued his doctoral studies at Inha University in South Korea (2025) under the guidance of Hyun-Chul Kim, focusing on hadron reactions and exotic states. He later joined the YST fellowship at APCTP, where his work centered on exotic pentaquarks~\cite{Clymton:2025dzt}. Similarly, Sakinah completed her master’s degree under the supervision of Terry Mart, and recently she just finished her doctoral degree on the topic of $J/\psi$ photoproduction at Kyungpook National University~\cite{Sakinah:2024cza} with Yongseok Oh and Tsung-Shung Harry Lee.
Muhammad Ridwan, who got a master's degree in UI~\cite{Ridwan:2024ngc}, pursued a doctoral degree in lattice QCD at Plymouth University, UK, under the supervision of Craig McNeile. 
These are among Terry Mart’s recent students who went abroad to pursue master’s and doctoral degrees.
The individuals mentioned above reflect only those known to the authors and may extend beyond what is presented here.
Overall, this shows that hadronic physics continues to grow in Indonesia, even though some researchers and students are currently based abroad.

\subsubsection{Active domestic research groups}
Currently active hadronic physics groups in Indonesia include the theoretical physics group at UI with Terry Mart, Anto Sulaksono, and colleagues, focusing on hyperons and neutron stars; the theoretical physics group at UGM with Chalis Setyadi and colleagues, focusing on gluon GTMD in the diffractive process; and experimental HEP group at BRIN, including Suharyo Sumowidagdo, Syaefudin Jaelani, Zulkaida Akbar, and others, with involvement in ALICE collaborations, as well as theoretical HEP group with Ardian N. Atmaja, Julio, Apriadi S. Adam, and colleagues, studying various HEP phenomenologies, including nonperturbative approaches. Meanwhile, other theoretical physics groups at several universities focus on different research areas.
Furthermore, BRIN and universities are actively collaborating, including through the mentoring of students through various programs, including an internship program.

\subsubsection{International Engagements}
In recent years, research collaborations have been done with several institutions abroad. In Japan, these include Tohoku University, RIKEN, J-PARC, Nagoya University, Osaka University, and Hiroshima University. In South Korea, collaborations involve Inha University, APCTP, Kyungpook National University, Soongsil University, Daegu University, Yonsei University, and Pukyong National University. In the US, there have been collaborations with George Washington University, Florida State University, Argonne National Laboratory, and North Carolina State University, while in Latin America, with the University of São Paulo, Universidade Federal de Goias (Brazil), and Universidad de Sonora (Mexico). In Europe, collaborations include those within the ALICE Collaboration, as well as with Utrecht University, Durham University, the University of Groningen, the University of Mainz, the University of Plymouth, and the University of Cambridge. In Australia, collaboration with Adelaide University.

Some Indonesian researchers have also participated in major international community conferences, including Hadron, Baryon, Quarks and Nuclear Physics (QNP), Hypernuclear, and Strange Particle Physics (HYP), Light-Cone (LC), Baryons, and Asia Pacific Few-Body (APFB) conferences, among others. In the past, only a small number of Indonesian researchers were involved in the international program activities.
Now, the international community has begun to recognize Indonesia’s presence in hadronic physics, but the number of active researchers remains relatively limited.

The International School of Strangeness Nuclear Physics (SNP), initiated by Osamu Hashimoto and his colleagues at Tohoku University, has played a significant role in fostering a collaborative community among young researchers in hadron and nuclear physics. It was later organized in partnership with J-PARC, further strengthening its academic and research activities. Since 2012, many of Terry Mart’s students have actively participated in this program, including the most recent school held in 2024.\footnote{\url{https://indico3.cns.s.u-tokyo.ac.jp/event/312/}}
Their in-person participation has largely been supported by the organizers of the SNP School. 

Furthermore, although Indonesia has long been a member of the ICTP, it only recently joined the APCTP in 2023. BRIN has played a leading role in establishing this partnership and continues to promote theoretical physics, including hadronic physics, through community building and international collaboration. To date, several researchers working in hadronic physics have been employed as postdoctoral fellows and as junior research group leaders.

\section{Present and Future Research Activities}
Here, we briefly mention several areas of hadronic physics that are currently being explored by Indonesian physicists and are expected to grow in the future.

\subsection{Hadron Reaction and Spectroscopy}
Research in hadron spectroscopy in Indonesia focuses on understanding the spectrum and internal dynamics of hadrons within the framework of QCD. Efforts include the investigation of missing nucleon resonances predicted by (constituent) quark models but not yet confirmed experimentally, the study of heavy baryons containing charm or bottom quarks to explore flavor dependence in baryon structure, and the search for exotic hadrons such as pentaquarks. Recently, many exotic hadrons have been detected in the experiment at CERN, which has become a great motivation to investigate this subject further theoretically.
These studies aim to deepen the understanding and knowledge of color confinement and the nonperturbative regime of QCD. Indonesian researchers have also contributed to international collaborations, such as the ALICE experiment at CERN, where they study quark-gluon plasma and hadron production, as well as their decay in heavy-ion collisions.

\subsection{Hadron Structure}
In the hadron structure studies, Indonesian researchers explore the internal structure of hadrons through the generalized parton distributions (GPDs), which are a combination of the electromagnetic form factors (EMFFs) and parton distribution functions (PDFs), parton distribution amplitudes (PDAs), transverse-momentum-dependent distributions (TMDs), and generalized transverse-momentum-dependent distributions (GTMDs) to provide a multidimensional picture (tomography) of the partonic structure of hadrons.
These studies are relevant and related to experimental activities in the EIC at BNL, the EicC in China, the COMPASS/AMBER++ at CERN, and JPARC in Japan. Another interesting subject of hadron structure is to explore the hadron structure in the medium, inspired by the structure function at finite nuclei. We have also investigated the hadron structure via the EMFFs, PDAs, in a nuclear medium. 
The results of this study will be useful and relevant for the ALICE, CMS, and FAIR experiments that measured the properties and hadron structure in medium. 

Very recently, Parada T. P. Hutauruk, in collaboration with Zulkaida Akbar, Apriadi S. Adam, and Chalis Setyadi, with their master student Taufiq I. Baihaqi, has developed a model-independent approach (data-driven model) using an artificial neural network (ANN) technique to predict the differential and total cross-section in the exclusive coherent diffractive $J/\psi$ production. 
Research direction in this topic will continue since the gluon distributions and dynamics will be measured at LHC, EIC, EicC, LHeC, and Glue-X at JLAB 22 GeV upgrade. Such a model-independent tool is useful to extract the physics information from the experimental data.

On the experimental side, 
since Indonesia's experimentalists are members of the ALICE collaboration and have been involved in the ALICE experiment, it is also good to continue discussing with them about what we can do from both sides to contribute to the international and national hadron research and community.

\subsection{Hyperon and Hypernuclear Physics}
\begin{figure}[t]
    \centering
    \includegraphics[width=0.7\textwidth]{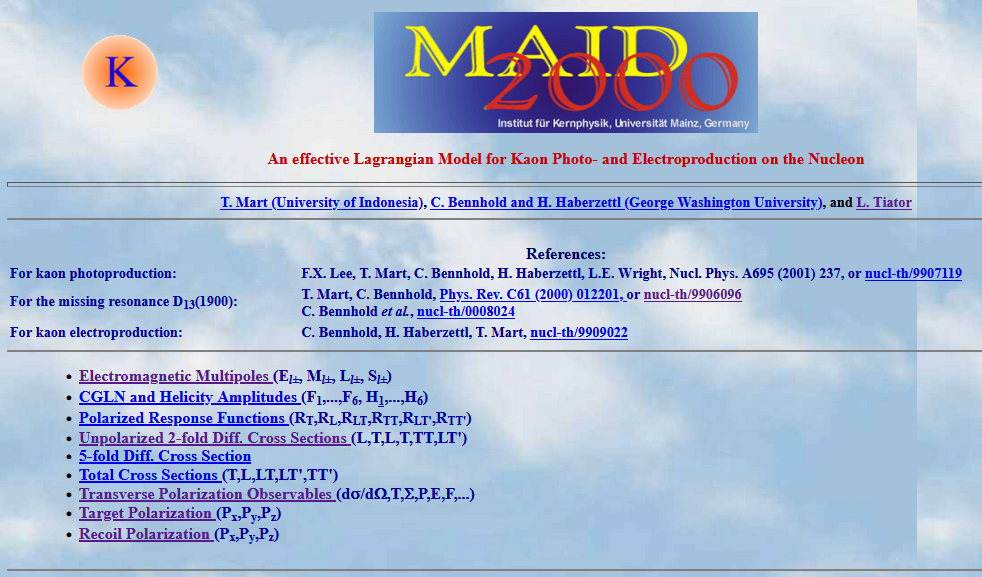}
    \caption{Kaon-Maid, an interactive model based on effective Lagrangian dynamics, designed to describe kaon photo- and electroproduction on the nucleon. The model is available at \url{https://maid.kph.uni-mainz.de/kaon/}.}
    \label{fig:kaon-maid}
\end{figure}

The production of hyperons is of particular interest because they are invariably accompanied by kaons, ensuring conservation of strangeness. For nearly three decades, Terry Mart has contributed extensively to this field, most notably through the development of a phenomenological model describing the electromagnetic production of kaons on the nucleon \cite{Mart:1999ed}. This model, known as Kaon-Maid\footnote{\url{https://maid.kph.uni-mainz.de/kaon/}}, was publicly released in 2000 via the Institut für Kernphysik at Universität Mainz (see Fig.~\ref{fig:kaon-maid}). Over the past 25 years, Kaon-Maid has become a valuable tool for both theorists and experimentalists, serving as a benchmark for theoretical predictions and as a generator for Monte Carlo simulations in experimental proposals. However, with the accumulation of extensive new measurements, the original formulation has become outdated, necessitating its revision. The elementary amplitude provided by this model remains crucial for theoretical studies of hypernuclear production, as electromagnetic reactions offer a particularly clean and well-understood mechanism \cite{Mart:1996ay,Mart:2008gq}.

\label{subsec:apfb08}

\subsection{Few-Body Physics}
Few-body physics is a branch of physics dealing with systems composed of a small number of interacting bodies. It therefore spans both classical and quantum regimes, but because many few-body studies focus on strong interactions among nucleons, the field overlaps substantially with hadronic physics.
At present, Imam Fachruddin at UI is the only researcher working actively in few-body physics. He also served as the chairman of APFB 2008 (see Subsec.~\ref{subsec:apfb08} for further details). His work primarily focuses on the formulation of few-nucleon systems using a three-dimensional operator approach~\cite{Golak:2010wz, Glockle:2010dfe}. Using this framework, he and his group have investigated kaon–nucleon scattering, as well as eta and kaon photoproduction on the nucleon. Another researcher working in few-body physics is Terry Mart, who collaborates with few-body specialist Emiko Hiyama at Tohoku University to investigate hypertriton binding energy and photoproduction using the Gaussian Expansion Method~\cite{Mart:2017mup}.

\subsection{Other Related Topics}
Theoretical frameworks developed in hadronic physics are also extended to the study of particle physics, dense nuclear systems, and astrophysical objects. In particular, hadron and quark models are employed to investigate the equation of state of matter in neutron stars, where modifications of hadron properties under extreme density and pressure provide valuable insights into the transition between hadronic and quark matter phases. Furthermore, the methodologies and effective field theories established in hadronic physics are closely linked to those used in particle physics beyond the Standard Model. These applications bridge hadronic physics with both nuclear astrophysics and particle physics, underscoring the interdisciplinary and unifying nature of the field.

\section{Community Activities}
The progress of hadronic physics in Indonesia has been closely linked to the growth of its research community. Here, we provide a general overview of the community’s activities, including 
conferences, and other initiatives that have contributed to the field’s development.

\subsection{General Conferences}
Several annual university and society conferences in Indonesia cover broad topics in mathematics and natural sciences, offering platforms for physicists to share and exchange ideas. University-based international conferences have become common since around 2015 and continue to expand nationwide. Notable examples include the International Symposium on Current Progress in Mathematics and Sciences (ISCPMS) organized by UI, and the Asian Physics Symposium (APS) hosted by ITB.
Meanwhile, the  Indonesian Conference on Theoretical and Applied Physics (ICTAP) connects Indonesian physicists with regional partners 
(see Sec.~\ref{subsec:ICTAP} for further information).

\subsection{Joint Conferences}
BRIN also plays an active role through joint workshops, such as the Conference on Accelerator-Based Science and Technology (CAST) 2024, chaired by Zulkaida Akbar and held jointly with the RIKEN Nishina Center, which strengthened cooperation in accelerator-based physics. BRIN has also initiated and organized joint workshops with APCTP and other international institutes, further reinforcing global partnerships in the field. More joint workshops are expected in the future, although these kinds of events often cover a broad range of topics.

\subsection{Local Meetings}
One of the major domestic conferences covering hadronic physics is the Conference on Theoretical Physics and Nonlinear Phenomena (CTPNP). It is primarily theory-oriented due to the limited number of experimentalists in Indonesia, and also sometimes organizes a general school on theoretical physics along with the conference. Although there is no school dedicated specifically to nuclear-hadronic physics, a particle physics school is held in Indonesia, supported by KEK, with the latest edition in 2025 at Universitas Islam Negeri Sunan Kalijaga, Yogyakarta.\footnote{\url{https://conference-indico.kek.jp/event/296/overview}}

\subsubsection{Conference on Theoretical Physics and Nonlinear Phenomena}
The CTPNP is an annual event organized by the Indonesian Theoretical Physicist Group (Grup Fisika Teoretik Indonesia, GFTI). Originating as a workshop of theoretical physics (WTP) in 2004 at UI, it developed into a full-scale conference in 2010.
Over the years, CTPNP has been hosted in major Indonesian cities, including Jakarta, Surabaya, Bandung, and Yogyakarta. In 2018, it was held outside Java for the first time, taking place in Makassar and organized by Hasanuddin University’s Department of Physics. 
Interestingly, Bum-Hoon Lee, who was the president of APCTP at that time, was actively involved in the recent conferences. 
The most recent meeting was held in Malang in 2019~\cite{CTPNP2019}. 

\subsubsection{International Conference on Theoretical and Applied Physics}
\label{subsec:ICTAP}
Despite not being exclusively focused on hadronic physics, the International Conference on Theoretical and Applied Physics (ICTAP), organized annually by the Physical Society of Indonesia (PSI)\footnote{\url{https://www.fisika.or.id/}} in collaboration with various universities across the country, provides an important forum for the field. The first meeting was held more than 14 years ago in Bandung, and the most recent one, the 14th ICTAP\footnote{\url{https://icsas.uns.ac.id}}, took place in Surakarta, Central Java, in 2025. The conference includes a dedicated session, where Indonesian hadron physicists present their research activities. 

\subsection{International Community Conferences}
Although Indonesia has several active physicists, it has hosted very few community-driven international conferences, and opportunities for large-scale gatherings that bring together researchers from across the region have so far been limited. 

\subsubsection{4th Asia-Pacific Conference on Few-Body Problems in Physics}
\label{subsec:apfb08}

\begin{figure}[t]
    \centering
    \includegraphics[width=0.7\textwidth]{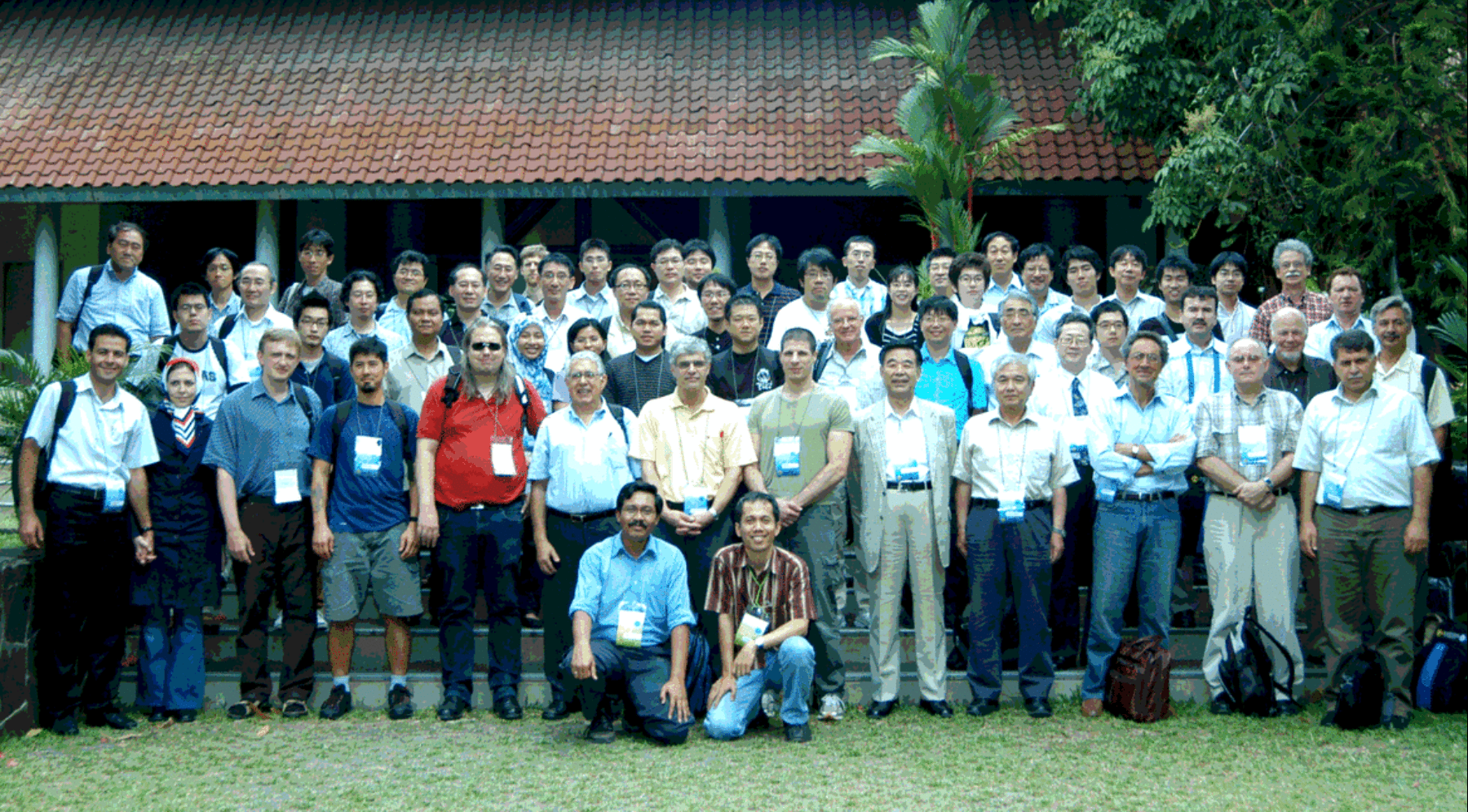}
    \caption{APFB 2008 at Universitas Indonesia, chaired by Imam Fachruddin and co-chaired by Terry Mart (seated in the front row).}
    \label{fig:apfb}
\end{figure}

The 4th Asia–Pacific Conference on Few-Body Problems in Physics (APFB08) was held at the University of Indonesia, Depok, from 19 to 23 August 2008. The conference brought together researchers from 19 countries across the Asia–Pacific region.
Although titled Asia-Pacific Conference, APFB08 also attracted participants from European countries: the United Kingdom, Germany, Portugal, the Netherlands, France, and Poland. This broad participation reflects the global nature of the few-body physics community, which organizes three major international conferences: the Asia–Pacific Conference on Few-Body Problems in Physics, the European Few-Body Conference, and the International Conference on Few-Body Problems in Physics. 
At APFB08, the largest delegations came from Japan (26 participants) and South Korea (11 participants), making it the largest international conference on this topic ever held in Indonesia. The scientific program covered a wide range of subjects, including hadronic and nuclear structures, reaction dynamics, and computational methods in few-body physics. 
The conference proceedings were published as a special issue of \textit{Modern Physics Letters A} in 2009~\cite{Fachruddin:2009zz}.
The APFB conference is held every three years; the most recent one (the 9th APFB or APFB2025) took place in Vietnam,\footnote{\url{https://indico.rcnp.osaka-u.ac.jp/event/2537/}} while the next meeting is scheduled in Busan, South Korea.

\begin{figure}[t]
    \centering
    \includegraphics[width=0.7\textwidth]{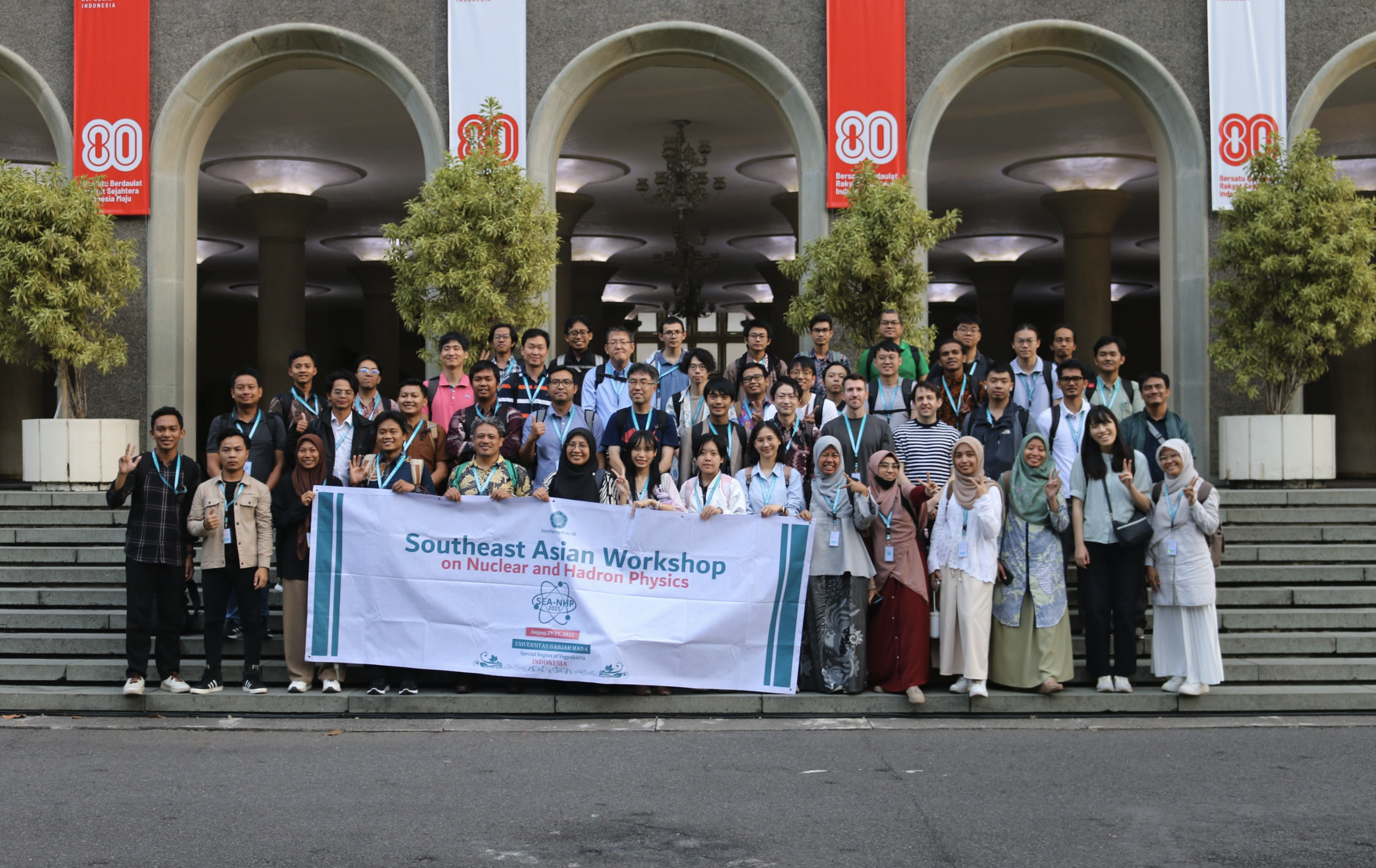}
    \caption{SEA-NHP 2025 in Universitas Gadjah Mada, chaired by Chalis Setyadi and co-chaired by Ahmad Jafar Arifi and Parada T. P. Hutauruk.}
    \label{fig:seanhp}
\end{figure}

\subsubsection{1st Southeast Asian Workshop on Nuclear and Hadronic Physics}
Building on these long-term efforts, Indonesia has participated in establishing the Southeast Asian Workshop on Nuclear and Hadronic Physics (SEA-NHP), a regional platform aimed at fostering collaboration among physicists in Southeast Asia. 
The SEA-NHP 2025 seeks to connect researchers working on nuclear and hadronic physics, encourage joint projects, and support the engagement of young scientists in international research activities. 
The inaugural workshop, held on 19-21 August 2025 at UGM, Yogyakarta,\footnote{\url{https://indico.global/event/13949/}} drew more than 50 participants from Indonesia, Thailand, Malaysia, and the Philippines, demonstrating the strong regional interest and active participation.
Participants from Japan and South Korea also attended in person, while others from Mexico, India, and Taiwan joined online.
These initiatives promote scientific exchange and collaboration in nuclear and hadronic physics across Southeast Asia. 
The 2nd SEA-NHP is scheduled to be held at Khon Kaen University, Thailand, and will be organized by Daris Samart as the chairman, continuing to expand regional collaboration.

\section{Challenges and Outlook}
In summary, the Indonesian hadronic physics community is steadily growing and becoming more active in international collaborations.
While it remains a relatively small community within the broader high-energy physics in Indonesia, its impact has been significant.
In the initial stages, most researchers pursued advanced-level studies in the US and Germany; however, more recently, they have expanded their efforts to other countries, such as Japan, the UK, and South Korea, reflecting a broader range of global engagement. 
Indonesian researchers have made relatively significant contributions in hadron reactions, spectroscopy, and structure research studies. These areas are expected to remain central to future research directions. 

Compared to 25 years ago, the academic and research environment and activities have experienced major improvement. Domestically, institutions such as UI, UGM, ITB, and BRIN have played a significant role in developing hadronic physics research and increasing the number of hadron physicists in Indonesia. BRIN has also taken a leading role through collaborations with university students. 
Engagement with international partners, such as ICTP and APCTP, has strengthened the research foundation and fostered collaborations, providing young researchers with global exposure. 
In recent years, APCTP has notably supported these efforts by offering fellowships and visiting programs. 

Looking ahead, the Indonesian hadronic physics community aims to enhance its research capabilities through sustained cooperation with institutions in the US, Europe, Japan, and South Korea. 
Moreover, although collaboration with researchers in China, Australia, and Latin America is still relatively limited in hadronic physics, it has been gradually increasing. 
With the growing enthusiasm of young scientists and increasing international visibility, Indonesia hopes to play a meaningful role in the next generation of research in hadronic physics, while continued support from the international community will further accelerate its development.

\section*{Acknowledgments}
We extend our gratitude to all colleagues who provided valuable insights and comments in the preparation of this manuscript. This work was supported by the PUTI Q1 Research Grant from the University of Indonesia (UI) under contract No. PKS-206/UN2.RST/HKP.05.00/2025. The works of A.J.A. and P.T.P.H. are supported by the RCNP Collaboration Research Network program under project number COREnet 057. 

\bibliographystyle{ws-mpla}
\bibliography{references}

@article{Gross:2022hyw,
  author       = {Gross, F. and Klempt, E. and Brodsky, S. J. and others},
  title        = {50 Years of Quantum Chromodynamics},
  journal      = {Eur. Phys. J. C},
  volume       = {83},
  pages        = {1125},
  year         = {2023},
  doi          = {10.1140/epjc/s10052-023-11949-2},
  url          = {https://doi.org/10.1140/epjc/s10052-023-11949-2}
}

@article{GlueX:2025hve,
    author = "Afzal, F. and others",
    collaboration = "GlueX",
    title = "{Measurement of the total compton scattering cross section between 6.5 and 11 GeV}",
    doi = "10.1016/j.physletb.2025.139914",
    journal = "Phys. Lett. B",
    volume = "870",
    pages = "139914",
    year = "2025"
}

@article{Gray:2024cof,
    author = "Gray, T. J. and others",
    title = "{Suppressed Electric Quadrupole Collectivity in $^{49}$Ti}",
    doi = "10.1016/j.physletb.2024.138856",
    journal = "Phys. Lett. B",
    volume = "855",
    pages = "138856",
    year = "2024"
}

@phdthesis{Silaban:1971,
  author       = {Pantur Silaban},
  title        = {Null Tetrad Formulation of the Equations of Motion in General Relativity},
  school       = {Syracuse University},
  year         = {1971},
  address      = {Syracuse, USA},
  type         = {Ph.D. Thesis}
}

@article{Pena:1996tf,
    author = "Pena, M. T. and Gross, Franz and Surya, Yohanes",
    title = "{Two pion exchange potential and the pi n amplitude}",
    doi = "10.1103/PhysRevC.54.2235",
    journal = "Phys. Rev. C",
    volume = "54",
    pages = "2235--2239",
    year = "1996"
}

@article{Gross:1992tj,
    author = "Gross, Franz and Surya, Yohanes",
    title = "{Unitary, relativistic resonance model for pi N scattering}",
    reportNumber = "CEBAF-TH-92-19, WM-92-110",
    doi = "10.1103/PhysRevC.47.703",
    journal = "Phys. Rev. C",
    volume = "47",
    pages = "703--723",
    year = "1993"
}

@article{Arifi:2020yfp,
    author = "Arifi, A. J. and Nagahiro, H. and Hosaka, A. and Tanida, K.",
    title = "{Roper-like resonances with various flavor contents and their two-pion emission decays}",
    doi = "10.1103/PhysRevD.101.111502",
    journal = "Phys. Rev. D",
    volume = "101",
    number = "11",
    pages = "111502",
    year = "2020"
}

@article{Ma:1981eg,
    author = "Ma, Ernest and Pramudita, A.",
    title = "{$K_L \to \gamma \gamma$ : Theory and Phenomenology}",
    reportNumber = "DOE-ER-00881-212",
    doi = "10.1103/PhysRevD.24.2476",
    journal = "Phys. Rev. D",
    volume = "24",
    pages = "2476",
    year = "1981"
}

@article{Mart:1995wu,
    author = "Mart, T. and Bennhold, C. and Hyde, Charles Earl",
    title = "{Constraints on coupling constants through charged sigma photoproduction}",
    reportNumber = "MKPH-T-94-26",
    doi = "10.1103/PhysRevC.51.R1074",
    journal = "Phys. Rev. C",
    volume = "51",
    pages = "R1074--R1077",
    year = "1995"
}

@article{Ji:1992xr,
    author = "Ji, C. R. and Surya, Y.",
    title = "{Calculation of scattering with the light cone two-body equation in phi**3 theories}",
    doi = "10.1103/PhysRevD.46.3565",
    journal = "Phys. Rev. D",
    volume = "46",
    pages = "3565--3575",
    year = "1992"
}

@article{D0:2004ruf,
    author = "Abazov, V. M. and others",
    collaboration = "D0",
    title = "{Measurement of dijet azimuthal decorrelations at central rapidities in $p\bar{p}$ collisions at $\sqrt{s} = 1.96$ TeV}",
    doi = "10.1103/PhysRevLett.94.221801",
    journal = "Phys. Rev. Lett.",
    volume = "94",
    pages = "221801",
    year = "2005"
}

@article{Sulaksono:2003re,
    author = "Sulaksono, A. and Burvenich, T. and Maruhn, J. A. and Reinhard, P. G. and Greiner, W.",
    title = "{The Nonrelativistic limit of the relativistic point coupling model}",
    doi = "10.1016/S0003-4916(03)00146-5",
    journal = "Annals Phys.",
    volume = "308",
    pages = "354--370",
    year = "2003"
}

@article{Hutauruk:2016sug,
    author = "Hutauruk, Parada T. P. and Cloet, Ian C. and Thomas, Anthony W.",
    title = "{Flavor dependence of the pion and kaon form factors and parton distribution functions}",
    doi = "10.1103/PhysRevC.94.035201",
    journal = "Phys. Rev. C",
    volume = "94",
    number = "3",
    pages = "035201",
    year = "2016"
}

@article{Salam:2004gz,
    author = "Salam, Agus and Arenhovel, Hartmuth",
    title = "{Interaction effects in K+ photoproduction on the deuteron}",
    doi = "10.1103/PhysRevC.70.044008",
    journal = "Phys. Rev. C",
    volume = "70",
    pages = "044008",
    year = "2004"
}

@article{Abidin:2009hr,
    author = "Abidin, Zainul and Carlson, Carl E.",
    title = "{Nucleon electromagnetic and gravitational form factors from holography}",
    doi = "10.1103/PhysRevD.79.115003",
    journal = "Phys. Rev. D",
    volume = "79",
    pages = "115003",
    year = "2009"
}

@article{ALICE:2017thy,
    author = "Acharya, Shreyasi and others",
    collaboration = "ALICE",
    title = "{$\Lambda_{\rm c}^+$ production in pp collisions at $\sqrt{s} = 7$ TeV and in p-Pb collisions at $\sqrt{s_{\rm NN}} = 5.02$ TeV}",
    reportNumber = "CERN-EP-2017-339",
    doi = "10.1007/JHEP04(2018)108",
    journal = "JHEP",
    volume = "04",
    pages = "108",
    year = "2018"
}

@article{Albright:1974nd,
    author = "Albright, Carl H. and Jarlskog, C. and Tjia, M. O.",
    title = "{Implications of Gauge Theories for Heavy Leptons}",
    reportNumber = "CERN-TH-1887",
    doi = "10.1016/0550-3213(75)90360-0",
    journal = "Nucl. Phys. B",
    volume = "86",
    pages = "535--547",
    year = "1975"
}

@article{Kusno:1980tf,
    author = "Kusno, Darmadi and Moravcsik, Michael J.",
    title = "{On the Problem of the Deuteron Smearing Corrections. 1. The Conventional Approach}",
    journal = "ICTP Report, IC-80-52",
    month = "",
    year = "1980"
}

@article{Badger:2017jhb,
    author = "Badger, Simon and Br{\o}nnum-Hansen, Christian and Hartanto, Heribertus Bayu and Peraro, Tiziano",
    title = "{First look at two-loop five-gluon scattering in QCD}",
    doi = "10.1103/PhysRevLett.120.092001",
    journal = "Phys. Rev. Lett.",
    volume = "120",
    number = "9",
    pages = "092001",
    year = "2018"
}

@article{CLAS:2017yjv,
    author = "Akbar, Z. and others",
    collaboration = "CLAS",
    title = "{Measurement of the helicity asymmetry $E$ in $\omega\to\pi^+\pi^-\pi^0$ photoproduction}",
    reportNumber = "JLAB-PHY-17-2532",
    doi = "10.1103/PhysRevC.96.065209",
    journal = "Phys. Rev. C",
    volume = "96",
    number = "6",
    pages = "065209",
    year = "2017"
}

@article{Barmawi:1966zz,
    author = "Barmawi, M.",
    title = "{Regge-Pole Analysis of pi++n --{\ensuremath{>}} omega+p}",
    doi = "10.1103/PhysRevLett.16.595",
    journal = "Phys. Rev. Lett.",
    volume = "16",
    pages = "595--597",
    year = "1966"
}

@article{Barmawi:1974yr,
    author = "Barmawi, M. and Venturi, Giovanni",
    title = "{A quark model for the pomeron}",
    doi = "10.1007/BF02762126",
    journal = "Lett. Nuovo Cim.",
    volume = "9S2",
    pages = "257--260",
    year = "1974"
}

@article{Barmawi:1968zz,
    author = "Barmawi, M.",
    title = "{Application of a Regge-Pole Model to the Reactions pi-p --{\ensuremath{>}} pi0n, pi-p --{\ensuremath{>}} etan, and pi+n --{\ensuremath{>}} omegap}",
    doi = "10.1103/PhysRev.166.1857",
    journal = "Phys. Rev.",
    volume = "166",
    pages = "1857--1862",
    year = "1968"
}

@article{Wospakrik:1982fd,
    author = "Wospakrik, H. J.",
    title = "{CLASSICAL EQUATION OF MOTION OF A SPINNING NONABELIAN TEST BODY IN GENERAL RELATIVITY}",
    doi = "10.1103/PhysRevD.26.523",
    journal = "Phys. Rev. D",
    volume = "26",
    pages = "523--526",
    year = "1982"
}

@article{Kusno:1978is,
    author = "Kusno, Darmadi and Moravcsik, Michael J.",
    title = "{The Existence of the West's Beta Correction}",
    reportNumber = "OITS-104",
    doi = "10.1103/PhysRevD.20.2734",
    journal = "Phys. Rev. D",
    volume = "20",
    pages = "2734",
    year = "1979"
}

@article{Albright:1966uhq,
    author = "Albright, C. H. and Liu, L. S. and Tjia, M. O.",
    title = "{Quark-Model Approach for the Semileptonic Reactions}",
    doi = "10.1103/PhysRevLett.16.921",
    journal = "Phys. Rev. Lett.",
    volume = "16",
    number = "20",
    pages = "921--926",
    year = "1966"
}

@article{Atmaja:2010uu,
    author = "Atmaja, Ardian Nata and de Boer, Jan and Shigemori, Masaki",
    title = "{Holographic Brownian Motion and Time Scales in Strongly Coupled Plasmas}",
    doi = "10.1016/j.nuclphysb.2013.12.018",
    journal = "Nucl. Phys. B",
    volume = "880",
    pages = "23--75",
    year = "2014"
}

@article{Ali:1996bm,
    author = "Ali, Ahmed and Hiller, G. and Handoko, L. T. and Morozumi, T.",
    title = "{Power corrections in the decay rate and distributions in $B \to X_s l^+ l^-$ in the Standard Model}",
    doi = "10.1103/PhysRevD.55.4105",
    journal = "Phys. Rev. D",
    volume = "55",
    pages = "4105--4128",
    year = "1997"
}

@article{Fachruddin:2000wv,
    author = "Fachruddin, I. and Elster, C. and Gloeckle, Walter",
    title = "{Nucleon-nucleon scattering in a three-dimensional approach}",
    doi = "10.1103/PhysRevC.62.044002",
    journal = "Phys. Rev. C",
    volume = "62",
    pages = "044002",
    year = "2000"
}

@article{Kiswandhi:2003ca,
    author = "Kiswandhi, Alvin and Capstick, Simon and Dytman, Steven",
    title = "{Model dependence of the properties of S(11) baryon resonances}",
    doi = "10.1103/PhysRevC.69.025205",
    journal = "Phys. Rev. C",
    volume = "69",
    pages = "025205",
    year = "2004"
}

@article{Hartanto:2013aha,
    author = "Hartanto, H. B. and Reina, L.",
    title = "{Hard-photon production with b jets at hadron colliders}",
    doi = "10.1103/PhysRevD.89.074001",
    journal = "Phys. Rev. D",
    volume = "89",
    number = "7",
    pages = "074001",
    year = "2014"
}

@article{Boer:2021upt,
    author = {Boer, Dani{\"e}l and Setyadi, Chalis},
    title = "{GTMD model predictions for diffractive dijet production at EIC}",
    doi = "10.1103/PhysRevD.104.074006",
    journal = "Phys. Rev. D",
    volume = "104",
    number = "7",
    pages = "074006",
    year = "2021"
}

@article{Boer:2023mip,
    author = {Boer, Dani{\"e}l and Setyadi, Chalis},
    title = "{Probing gluon GTMDs through exclusive coherent diffractive processes}",
    doi = "10.1140/epjc/s10052-023-12040-6",
    journal = "Eur. Phys. J. C",
    volume = "83",
    number = "10",
    pages = "890",
    year = "2023"
}

@article{Muzakka:2022wey,
    author = "Muzakka, K. F. and others",
    title = "{Compatibility of neutrino DIS data and its impact on nuclear parton distribution functions}",
    doi = "10.1103/PhysRevD.106.074004",
    journal = "Phys. Rev. D",
    volume = "106",
    number = "7",
    pages = "074004",
    year = "2022"
}

@article{Kiswandhi:2011cq,
    author = "Kiswandhi, Alvin and Yang, Shin Nan",
    title = "{On the near-threshold peak structure in the differential cross section of {\textbackslash}phi-meson photoproduction: indication of a missing resonance with non-negligible strangeness content}",
    doi = "10.1103/PhysRevC.86.015203",
    journal = "Phys. Rev. C",
    volume = "86",
    pages = "015203",
    year = "2012",
}

@article{Arifi:2022pal,
    author = "Arifi, Ahmad Jafar and Choi, Ho-Meoyng and ji, Chueng-Ryong and Oh, Yongseok",
    title = "{Mixing effects on 1S and 2S state heavy mesons in the light-front quark model}",
    doi = "10.1103/PhysRevD.106.014009",
    journal = "Phys. Rev. D",
    volume = "106",
    number = "1",
    pages = "014009",
    year = "2022"
}

@article{Ridwan:2024ngc,
    author = "Ridwan, Muhammad and Arifi, Ahmad Jafar and Mart, Terry",
    title = "{Self-consistent M1 radiative transitions of excited Bc and heavy quarkonia with different polarizations in the light-front quark model}",
    doi = "10.1103/PhysRevD.111.016011",
    journal = "Phys. Rev. D",
    volume = "111",
    number = "1",
    pages = "016011",
    year = "2025"
}

@article{Sakinah:2024cza,
    author = "Sakinah, S. and Lee, T. -S. H. and Choi, Ho-Meoyng",
    title = "{Dynamical model of J/{\ensuremath{\psi}} photoproduction on the nucleon}",
    doi = "10.1103/PhysRevC.109.065204",
    journal = "Phys. Rev. C",
    volume = "109",
    number = "6",
    pages = "065204",
    year = "2024"
}

@inproceedings{CTPNP2019,
  author       = {Alatas, H and others},
  title        = {Preface: Conference on Theoretical Physics and Nonlinear Phenomena (CTPNP) 2019},
  booktitle    = {AIP Conference Proceedings},
  volume       = {2234},
  pages        = {010001},
  year         = {2020},
  doi          = {10.1063/12.0000525},
  url          = {https://doi.org/10.1063/12.0000525}
}

@proceedings{Fachruddin:2009zz,
    editor = "Fachruddin, Imam and Mart, Terry",
    title = "{Few-body problems in physics. Proceedings, 4th Asia-Pacific Conference, AFFB'08, Depok, Indonesia, August 19-23, 2008; Mod. Phys. Lett. A }",
    journal = {Mod. Phys. Lett. A},
    volume = "24",
    pages = "pp.779--1086",
    year = "{\bf 24}, 779--1086 (2009)"
}

@article{Mart:1996ay,
    author = "Mart, T. and Tiator, L. and Drechsel, D. and Bennhold, C.",
    title = "{Electromagnetic production of the hypertriton}",
    doi = "10.1016/S0375-9474(98)00441-2",
    journal = "Nucl. Phys. A",
    volume = "640",
    pages = "235--258",
    year = "1998"
}

@article{Mart:2008gq,
    author = "Mart, T. and Van Der Ventel, Brandon",
    title = "{Photo- and Electroproduction of the Hypertriton on He-3}",
    doi = "10.1103/PhysRevC.78.014004",
    journal = "Phys. Rev. C",
    volume = "78",
    pages = "014004",
    year = "2008"
}

@article{Mart:1999ed,
    author = "Mart, T. and Bennhold, C.",
    title = "{Evidence for a missing nucleon resonance in kaon photoproduction}",
    doi = "10.1103/PhysRevC.61.012201",
    journal = "Phys. Rev. C",
    volume = "61",
    pages = "012201",
    year = "1999"
}

@article{Hutauruk:2018zfk,
    author = {Hutauruk, Parada T. P. and Bentz, Wolfgang and Clo{\"e}t, Ian C. and Thomas, Anthony W.},
    title = "{Charge Symmetry Breaking Effects in Pion and Kaon Structure}",
    doi = "10.1103/PhysRevC.97.055210",
    journal = "Phys. Rev. C",
    volume = "97",
    number = "5",
    pages = "055210",
    year = "2018"
}

@article{Abidin:2019xwu,
    author = "Abidin, Zainul and Hutauruk, Parada T. P.",
    title = "{Kaon form factor in holographic QCD}",
    doi = "10.1103/PhysRevD.100.054026",
    journal = "Phys. Rev. D",
    volume = "100",
    number = "5",
    pages = "054026",
    year = "2019"
}

@article{Kristiano:2017qjq,
    author = "Kristiano, J. and Clymton, S. and Mart, T.",
    title = "{Pure spin-3/2 propagator for use in particle and nuclear physics}",
    doi = "10.1103/PhysRevC.96.052201",
    journal = "Phys. Rev. C",
    volume = "96",
    number = "5",
    pages = "052201",
    year = "2017"
}

@article{Kristiano:2022maq,
    author = "Kristiano, Jason and Yokoyama, Jun'ichi",
    title = "{Constraining Primordial Black Hole Formation from Single-Field Inflation}",
    reportNumber = "RESCEU-20/22",
    doi = "10.1103/PhysRevLett.132.221003",
    journal = "Phys. Rev. Lett.",
    volume = "132",
    number = "22",
    pages = "221003",
    year = "2024"
}

@article{Golak:2010wz,
    author = "Golak, J. and Glockle, W. and Skibinski, R. and Witala, H. and Rozpedzik, D. and Topolnicki, K. and Fachruddin, I. and Elster, Ch. and Nogga, A.",
    title = "{The Two-Nucleon System in Three Dimensions}",
    doi = "10.1103/PhysRevC.81.034006",
    journal = "Phys. Rev. C",
    volume = "81",
    pages = "034006",
    year = "2010"
}

@article{Glockle:2010dfe,
    author = "Glockle, W. and Fachruddin, I. and Elster, Ch. and Golak, J. and Skibinski, R. and Witala, H.",
    title = "{3N Scattering in a Three-Dimensional Operator Formulation}",
    doi = "10.1140/epja/i2010-10920-4",
    journal = "Eur. Phys. J. A",
    volume = "43",
    pages = "339--350",
    year = "2010"
}

@article{Mart:2017mup,
    author = "Mart, Terry and Hiyama, Emiko",
    editor = "Tamura, H.",
    title = "{Photoproduction of Hyperon and Hypertriton at Forward Angles}",
    doi = "10.7566/JPSCP.17.062004",
    journal = "JPS Conf. Proc.",
    volume = "17",
    pages = "062004",
    year = "2017"
}

@article{AlGhifari:2025xrt,
    author = "Al Ghifari, M. H. and Ramadhan, H. S. and Alatas, H. and Sulaksono, A.",
    title = "{Bound on generalized uncertainty principle parameter from nuclear matter and slow rotating neutron stars}",
    doi = "10.1007/s10714-025-03457-3",
    journal = "Gen. Rel. Grav.",
    volume = "57",
    number = "8",
    pages = "121",
    year = "2025"
}

@article{CMS:2012qbp,
    author = "Chatrchyan, Serguei and others",
    collaboration = "CMS",
    title = "{Observation of a New Boson at a Mass of 125 GeV with the CMS Experiment at the LHC}",
    reportNumber = "CMS-HIG-12-028, CERN-PH-EP-2012-220",
    doi = "10.1016/j.physletb.2012.08.021",
    journal = "Phys. Lett. B",
    volume = "716",
    pages = "30--61",
    year = "2012"
}

@article{ALICE:2022wpn,
    author = "Acharya, Shreyasi and others",
    collaboration = "ALICE",
    title = "{The ALICE experiment: a journey through QCD}",
    doi = "10.1140/epjc/s10052-024-12935-y",
    journal = "Eur. Phys. J. C",
    volume = "84",
    number = "8",
    pages = "813",
    year = "2024"
}

@article{Clymton:2021wof,
    author = "Clymton, S. and Mart, T.",
    title = "{Extracting the pole and Breit-Wigner properties of nucleon and {\ensuremath{\Delta}} resonances from the {\ensuremath{\gamma}}N{\textrightarrow}K{\ensuremath{\Sigma}} photoproduction}",
    doi = "10.1103/PhysRevD.104.056015",
    journal = "Phys. Rev. D",
    volume = "104",
    number = "5",
    pages = "056015",
    year = "2021"
}

@article{Clymton:2025dzt,
    author = "Clymton, Samson and Kim, Hyun-Chul and Mart, Terry",
    title = "{Triple-strangeness hidden-charm pentaquarks}",
    doi = "10.1103/6p7j-kl11",
    journal = "Phys. Rev. D",
    volume = "112",
    number = "9",
    pages = "094024",
    year = "2025"
}

@article{Hutauruk:2018cgu,
    author = "Hutauruk, Parada T. P. and Oh, Yongseok and Tsushima, K.",
    title = "{Impact of medium modifications of the nucleon weak and electromagnetic form factors on the neutrino mean free path in dense matter}",
    doi = "10.1103/PhysRevD.98.013009",
    journal = "Phys. Rev. D",
    volume = "98",
    number = "1",
    pages = "013009",
    year = "2018"
}

@article{Hutauruk:2018qku,
    author = "Hutauruk, Parada T. P. and Oh, Yongseok and Tsushima, K.",
    title = "{Electroweak properties of pions in a nuclear medium}",
    doi = "10.1103/PhysRevC.99.015202",
    journal = "Phys. Rev. C",
    volume = "99",
    number = "1",
    pages = "015202",
    year = "2019"
}

@article{Arifi:2023jfe,
    author = "Arifi, Ahmad Jafar and Hutauruk, Parada. T. P. and Tsushima, Kazuo",
    title = "{In-medium properties of the light and heavy-light mesons in a light-front quark model}",
    doi = "10.1103/PhysRevD.107.114010",
    journal = "Phys. Rev. D",
    volume = "107",
    number = "11",
    pages = "114010",
    year = "2023"
}

@article{Kim:2013cpa,
    author = "Kim, C. S. and Li, Run-Hui and Simanjuntak, Freddy and Zou, Z. T.",
    title = "{Charmless $B_{u,d,s} \to VT$ decays in perturbative QCD approach}",
    doi = "10.1103/PhysRevD.88.014031",
    journal = "Phys. Rev. D",
    volume = "88",
    number = "1",
    pages = "014031",
    year = "2013"
}

@phdthesis{Wiranata:2011kwb,
    author = "Wiranata, Anton",
    title = "{Transport Coefficients of Interacting Hadrons}",
    school = "Ohio University",
    year = "2011"
}

@article{Mart:2013gfa,
    author = "Mart, T. and Sulaksono, A.",
    title = "{Nonidentical protons}",
    doi = "10.1103/PhysRevC.87.025807",
    journal = "Phys. Rev. C",
    volume = "87",
    number = "2",
    pages = "025807",
    year = "2013"
}

@article{Suparti:2017msx,
    author = "Suparti and Sulaksono, A. and Mart, T.",
    title = "{Influence of the nucleon radius on the properties of slowly rotating neutron stars}",
    doi = "10.1103/PhysRevC.95.045806",
    journal = "Phys. Rev. C",
    volume = "95",
    number = "4",
    pages = "045806",
    year = "2017"
}

@phdthesis{Riveli:2014fpb,
    author = "Riveli, Nowo",
    title = "{Direct Photon - Hadron Correlations Measurement in Au+Au Collision at Nucleon
Center-Of-Mass Energy of 200 GeV With Isolation Cut Methods}",
    school = "Ohio University",
    year = "2014"
}

@article{Carlson:2003xb,
    author = "Carlson, Carl E. and Carone, Christopher D. and Kwee, Herry J. and Nazaryan, Vahagn",
    title = "{A Naturally narrow positive parity Theta+}",
    doi = "10.1103/PhysRevD.70.037501",
    journal = "Phys. Rev. D",
    volume = "70",
    pages = "037501",
    year = "2004"
}

@article{Kwee_2008,
   title={Finite-Size Correction in Many-Body Electronic Structure Calculations},
   volume={100},
   ISSN={1079-7114},
   url={http://dx.doi.org/10.1103/PhysRevLett.100.126404},
   DOI={10.1103/physrevlett.100.126404},
   number={12},
   journal={Physical Review Letters},
   publisher={American Physical Society (APS)},
   author={Kwee, Hendra and Zhang, Shiwei and Krakauer, Henry},
   year={2008},
   month=mar }

@article{Carone:2006wj,
    author = "Carone, Christopher D. and Erlich, Joshua and Tan, Jong Anly",
    title = "{Holographic Bosonic Technicolor}",
    doi = "10.1103/PhysRevD.75.075005",
    journal = "Phys. Rev. D",
    volume = "75",
    pages = "075005",
    year = "2007"
}
\end{document}